\documentclass[5p,sort&compress]{elsarticle}
\usepackage{amssymb}
\usepackage{amsmath}
\usepackage{color}
\usepackage{graphicx}
\usepackage{caption}
\usepackage{subcaption}
\usepackage[english]{babel}

\newcommand{\er}{$\pm$}

\begin{document}

\begin{frontmatter}

\title{$N^*\to N \eta^\prime$ decays from photoproduction of $\eta^\prime$-mesons off protons}

\author[label1,label2]{A.~V.~Anisovich}
\author[label3]{V.~Burkert}
\author[label4]{P.~M.~Collins}
\author[label4]{M.~Dugger}
\author[label1,label3]{E.~Klempt}
\author[label1,label2]{V.~A.~Nikonov}
\author[label4]{B.~G.~Ritchie}
\author[label1,label2]{A.~Sarantsev}
\author[label1]{U.~Thoma}

\address[label1]{Helmholtz--Institut f\"ur Strahlen-- und Kernphysik, Universit\"at Bonn, Germany}
\address[label2]{NRC ``Kurchatov Institute'', PNPI, Gatchina 188300, Russia}
\address[label3]{Thomas Jefferson National Accelerator Facility, Newport News, Virginia 23606}
\address[label4]{Arizona State University, Tempe, Arizona 85287-1504}

\begin{abstract}
A study of the partial-wave content of the $\gamma p\to \eta^\prime p$ reaction in the
fourth resonance region is presented, which has been prompted by new measurements of
polarization observables for that process. Using the Bonn-Gatchina partial-wave formalism,
the incorporation of new data indicates that the $N(1895)1/2^-$, $N(1900)3/2^+$,
$N(2100)1/2^+$, and $N(2120)3/2^-$ are the most significant contributors to the
photoproduction process. New results for the branching ratios of the decays of these
more prominent resonances to $N\eta^\prime$ final states are provided; such branches
have not been indicated in the most recent edition of the Review of Particle
Properties. Based on the analysis performed here, predictions for the helicity
asymmetry $E$ for the $\gamma p\to \eta^\prime p$ reaction are presented.
 \end{abstract}


\end{frontmatter}
\section{Introduction}
The cross section for pion-nucleon elastic scattering as a function of center-of-mass
energy $W$ reveals four distinct but broad energy ranges where enhancements
are observed, which are called resonance regions. The first resonance region is
principally due to $\Delta(1232)3/2^+$ formation, which dominates the cross section
at low masses. Somewhat higher in $W$, the second resonance region houses the
$N(1520)3/2^-$ as the leading resonance, along with contributions from the
$N(1440)1/2^+$ and $N(1535)1/2^-$ excitations. At still higher $W$, several well-known
resonances contribute to the third resonance region, in particular the $N(1680)5/2^+$
state. At $1900\leq W\leq 2100$\,MeV, the fourth resonance region appears as a
small peak-like structure in the total $\pi N$ cross section, which is largely due
to the $\Delta(1950)$ $7/2^+$ excitation with substantial contributions from other
$\Delta^*$ resonances.

Interestingly, $N^*$ resonance contributions to the fourth resonance region have been
difficult to identify, and these contributions are presently under study in a number
of experiments. Photoproduction of $\eta^\prime$-mesons offers the chance to search
for low-spin high-mass nucleon resonances in the region above $W=1900$ MeV. Due to
isospin conservation, the reaction $\gamma p\to \eta^\prime p$ receives contributions
only from $N^*$ intermediate states, and thus the reaction functions as an ``isospin
filter" for the nucleon resonance spectrum, helping isolate those $N^*$ states.
Furthermore, due to the angular momentum barrier, high-spin resonances are suppressed.
Consequently, the photoproduction of $\eta^\prime$ mesons can be expected to shed
light on the low-spin $N^*$ resonances in the fourth resonance region.

Photoproduction of  $\eta^\prime$-mesons was first studied at DESY
in a hydrogen bubble chamber \cite{ABBHHM:1968aa} and a streamer
chamber \cite{Struczinski:1975ik}, though only a few events were
observed for incident photon energies up to 6.3\,GeV. At ELSA in
Bonn, the reaction was investigated with the magnetic spectrometer
SAPHIR in the energy range from 900 to 2600\,MeV; 250 events due to
$\eta^\prime$ production were reported. The linear forward rise of
the angular distribution was assigned to two resonances with
$J^P=1/2^\pm$ \cite{Plotzke:1998ua}. Within the MAID model, the
SAPHIR data were described by the interference of a $J^P=1/2^-$
resonance and the exchange of a $t$-channel Regge trajectory
\cite{Chiang:2002vq}. When the new data from CLAS on $\eta^\prime$
photoproduction on the proton for 1935\,MeV $<W<$
2249\,MeV~\cite{Dugger:2005my} were included in the fit, four
resonances with $J^P=1/2^\pm$ and $J^P=3/2^\pm$ were required to
achieve a good description~\cite{Tiator:2006he} of the data. Huang,
Haberzettl, and Nakayama~\cite{Huang:2012xj} included additionally
data on the differential cross sections on $\eta^\prime$
photoproduction off protons from the CLAS collaboration, which
covered the range from the production threshold up to
$W=2840$\,MeV~\cite{Williams:2009yj}, and the CBELSA/TAPS data off
protons~\cite{Crede:2009zzb} and off nucleons bound in the
deuteron~\cite{Jaegle:2010jg} taken from the production threshold up
to $W=2380$\,MeV. The fit required a subthreshold contribution of
$N(1720)3/2^+$ and three above-threshold resonances with quantum
numbers $J^P=1/2^\pm$ and $3/2^+$. Recently, the A2 Collaboration at
MAMI studied the reaction $\gamma p\to \eta p$ and $\to
\eta'p$~\cite{Kashevarov:2017kqb}. A strong cusp is observed in the
$\eta$ excitation functions in vicinity of the $\eta'$ threshold.
Within the $\eta$-MAID isobar model, the cusp is interpreted by
production of the $N(1895)1/2^-$ nucleon resonance and its
significant decay branching ratios to both, to $N\eta$ and $N\eta'$.
Indeed, all analyses agree that the threshold in $\eta^\prime$
photoproduction is governed by a resonance with $J^P=1/2^-$ and a
mass of about 1900\,MeV. In addition, there is evidence for
$N(2100)1/2^+$, and contributions from  a $3/2^+$ resonance from a
$3/2^-$ resonance also have been suggested.

Partial-wave analyses benefit tremendously from data on meson photoproduction observables
obtained when different relative orientations of the spin of the incident photon or the
struck nucleon (or both) are available (so-called ``polarization observables"). Such data are
now becoming available for the $\gamma p\to \eta^\prime p$ reaction. Recently, the GRAAL
collaboration reported a measurement of the linearly-polar\-ized photon beam asymmetry
$\Sigma$ for $\eta^{\prime}$ photoproduction on unpolarized protons near
threshold~\cite{Sandri:2014nqz}. More recently, the CLAS collaboration~\cite{Collins:2017sgu}
also measured $\Sigma$ over an extended mass range from threshold up to 2092\,MeV.
These new data on $\Sigma$ stimulated us to study the partial-wave content of the
$\gamma p\to \eta^\prime p$ reaction using a model that also describes simultaneously
data for the photoproduction of other mesons as well. Such multi-channel analyses can
provide great insight into the nucleon resonance spectrum since the strengths of the
decay modes for the various resonances participating in the process vary greatly from
one final state to another.

\section{Formalism}
For the investigation reported here, a partial-wave analysis was performed with the
Bonn-Gatchina (BnGa) formalism, described more fully
elsewhere~\cite{Anisovich:2004zz,Anisovich:2006bc,Anisovich:2007zz,Denisenko:2016ugz}.
Briefly, this approach uses a modified $K$-matrix in the form
 \begin{equation}
 \mathbf{\hat A}(s) \;=\; \mathbf{\hat K}\;(\mathbf{\hat I}\;-
 \mathbf{\hat B \hat K})^{-1} \label{k_matrix}
\end{equation}
that is defined in the complex $s$ plane. On the real axis, $\sqrt s=W$.
The $K$-matrix elements combine the contributions from resonances and from background through
\begin{equation}
\label{kmat}
K_{ab}=\sum\limits_\alpha \frac{g^{\alpha}_a
g^{\alpha}_b}{M_\alpha^2-s}+f_{ab}\,.
\end{equation}
Here, $g^{\alpha}_{a,b}$ are coupling constants of the pole $\alpha$ to the initial state
$a$ and the final state $b$. $M_\alpha$ defines the $K$-matrix pole (which differs from
the $T$-matrix pole which is listed in the RPP). $f_{ab}$ represents non-resonant
background contributions.
$f_{ab}$ could be functions of $s$ but in practice, a constant term is sufficient
to achieve good fits. Only for the $J^P=1/2^-$ wave, a more complicated expression
was used:
\begin{equation}
\label{f-swave}
f_{ab}(s)= \frac{f_{ab}^{(1)}+f_{ab}^{(2)}\sqrt s}{s-s_0^{ab}}
\end{equation}
where $f_{ab}^{(1)}, f_{ab}^{(2)}, s_0^{ab}$ are constants determined in the fit.

The multi-channel amplitude {\boldmath$\hat{{\rm A}}(s)$} with
matrix elements ${A}_{ab}(s)$ defines the transition amplitude from channel $a$
to channel $b$. $\mathbf{\hat B}$ is a diagonal matrix with an imaginary
part given by the corresponding phase space volume
\begin{equation}
\hat B_i= \Re e B_i+\;i\rho_i\,,
\end{equation}
where $\Re e B_i$ is calculated from the dispersion integral with one subtraction
regularization. In addition to this modified $K$-matrix, we have also included a Regge-ized amplitude
describing the exchange of vector mesons in the $t$-channel~\cite{Denisenko:2016ugz}.
For the fits of the existing data with this approach, we restricted the mass range to
$W\leq 2360$\,MeV since many nucleon resonances will contribute to the process at high
incident photon energies, making it difficult to identify the leading contributions.

In the reaction $\gamma p\to \eta'p$, the coupling constants in the
initial and the final state are weak. In this case, a simplified
amplitude can be used
\begin{eqnarray}
\hspace{-6mm} A_{f}^h &=& \hat G_{f} + \hat P_{a}[(\hat I\;-\;\hat B
\hat K)^{-1}\,\hat B ]_{ab} \hat D_{bf}\,.
 \label{PDvect}
 \end{eqnarray}
The transition from the initial $\gamma N$ state to the K-matrix
channel $a$ is represented by
\begin{eqnarray}
\hspace{-5mm}\hat P_a&=&\sum\limits_\alpha \frac{g_{\gamma
N}^{(\alpha)} g^{\alpha}_a}{M_\alpha^2-s}+F_{a}
\end{eqnarray}
where $g_{\gamma N}^{(\alpha)}$ are resonance couplings and $F_a$
describes the non-resonant transition.
\begin{eqnarray}
\hat D_{bf} &=& \sum_\alpha \frac{g_b^{(\alpha)}g_{f}^{(\alpha)}
}{M^2_\alpha - s} \;+\;
 \tilde d_{bf}\
\end{eqnarray}
represents the transition from the channel $b$ to the final
state '$f$'. $\hat G_{f}$ corresponds to a tree diagram for the
transition from initial channel ($\gamma N$ in the case
photoproduction)  to the state '$f$':
\begin{eqnarray}
\hat  G_{f}\;=\;\sum_\alpha \frac{g_{\gamma N}^{(\alpha)}g_{f}^{(\alpha)} }{M^2_\alpha - s} \;+\;
 \tilde h_{(\gamma N)f}\,.
\end{eqnarray}
Direct non-resonant transitions to the $\eta'N$ final state $\tilde
d_{bf}$ and $\tilde h_{(\gamma N)f}$ were set to zero in this
analysis and only the decay couplings $g_f^{(\alpha)}$ were fitted as
free parameters.

The forward peak in the differential cross section at high energies
is described by a Reggeized $\rho(\omega)$-meson exchange amplitude
\cite{Anisovich:2004zz,Sarantsev:2008ar}. If the $\eta'N$
photoproduction data are fitted with Eqn.~(\ref{PDvect}) the
projection of the t,u-channel amplitude into partial waves only
contributes to the $\tilde h_{(\gamma N)f}$ term. If the $\eta'N$
channel is taken into account as a channel, the
t,u-channel amplitudes should contribute also to the P-vectors and
to the full amplitude due to rescattering via a $\eta' N$ meson loop
diagram. However, such a contribution is very small compare to the
other components (e.g. nonresonant transition to $\pi N$ channel)
and can be neglected in the fit.

With $g(t)=g_0\exp(-bt)$ as vertex function and form factor,
$\alpha(t)$ as trajectory, $\nu=\frac 12 (s-u)$, $\nu_0$
as normalization factor, we obtain
\cite{Anisovich:2004zz}:
\begin{eqnarray}
\label{eq9a}
A&=&g(t)\cdot\frac{e^{-i\frac{\pi}{2}\alpha(t)}} {\cos
(\frac{\pi}{2}\alpha(t))} \left (\frac{\nu}{\nu_0}\right
)^{\alpha(t)}\;.
\end{eqnarray}
To eliminate the poles at $t<0$, additional $\Gamma$-functions are
introduced in (\ref{eq9a}) by replacing
\begin{eqnarray}
\cos \left (\frac{\pi}{2}\alpha(t)\right ) \to \cos \left
(\frac{\pi}{2}\alpha(t)\right )  \; \Gamma \left (\frac
{\alpha(t)}{2} +\frac12\right )\, .
\label{rho_1}
\end{eqnarray}

The $\rho(\omega)$ trajectory is parametrized as
\begin{eqnarray}
\label{eq11a}
\hspace{-5mm}\rho(\omega)\quad\qquad\qquad\alpha(t)&=&0.50+0.85
({\rm GeV}^{-2})t
\end{eqnarray}
where $t$ is given in GeV$^2$. The amplitude $A$ in Eq.~(\ref{eq9a})
is related to the cross section by
\begin{eqnarray}
\label{eq12}
\frac{d\sigma}{dt}=\frac{1}{64\pi s k^2}|A|^2
\end{eqnarray}

\section{Fits to the data on $\gamma p\to\eta^\prime p$}

In addition to the data on $\eta'$ production, we used further data
to constrain the amplitudes. We used the results on the real and imaginary part of
the $\pi N$ scattering amplitude for the partial waves up to
$J=9/2^\pm$ from Ref.~\cite{Arndt:2006bf} and data on pion and
photoproduction data with $\pi N$, $\eta N$, $K\,\Lambda$,
$K\,\Sigma$, $N\pi^0\pi^0$, and $N\pi^0\eta$ in the final state. A
list of the data with references can be found on our web page
(pwa.hiskp.uni-bonn.de/). This primary fit was used for a study of
$N^*\to N\omega$~\cite{Denisenko:2016ugz}, $\to
N\eta$~\cite{Muller:2017sub}, $\to
K^*\Lambda$~\cite{Anisovich:2017sub}, and now for $N\eta'$ decays.

The primary fits for this analysis used all data listed above except those on
$\gamma p\to p\omega$ and $\gamma p\to K^{*+}\Lambda$. Different primary fits were made,
all gave a good description of the data. The fits differed in the number of  $K$-matrix poles
assumed to contribute to the $J^P=3/2^+$ partial wave (corresponding to
resonances at 1720, 1900, 2200, with or without
a resonance at 1960\,MeV), two or three resonances in the $J^P=5/2^+$ partial wave
(at 1680 and 1860 and/or 2100\,MeV), different helicity couplings for $N(1990)7/2^+$.
Furthermore, resonances above 2\,GeV with spin parities up to $J=7/2$
were added (one by one) in all partial waves. Sometimes a better fit was
achieved but the gain in $\chi^2$ was
not sufficient to claim evidence for new resonances.

For most fits which included the data on $\gamma p\to\eta^\prime p$,
all parameters of one of these fits were frozen, and only additional
$N\eta'$ decay were admitted.  These fits used different primary
fits and resulted in a variation of $N^*\to N\eta'$ branching
ratios. In final fits, all parameters were varied but they remained
stably in the local minimum.

For reaction $\gamma p\to\eta^\prime p$, the following data were
included: The differential cross sections from the CBELSA/\ TAPS
(CBT) experiment~\cite{Crede:2009zzb} cover the range from threshold
to 2360\,MeV in invariant mass. The data are divided into 20 mass
bins and 10 bins in $\cos\theta$, where $\theta$ gives the
$\eta^\prime$ direction in the center-of-mass system. The more
recent CLAS publication on the process~\cite{Williams:2009yj}
reported differential cross sections up to $W$=2840\,MeV, in 40 mass
bins and 20 angular bins with superior statistics. The existence of
two different data sets with different systematics allows for
valuable cross checks on the results from the two experiments.

The CLAS differential cross sections have a systematic uncertainty of about 12\%, which,
when used as a scaling factor in the fit, helped increase the compatibility of the
\begin{figure}[h!]
 \center
  \begin{subfigure}[t]{0.48\textwidth}
    \includegraphics[width=\textwidth]{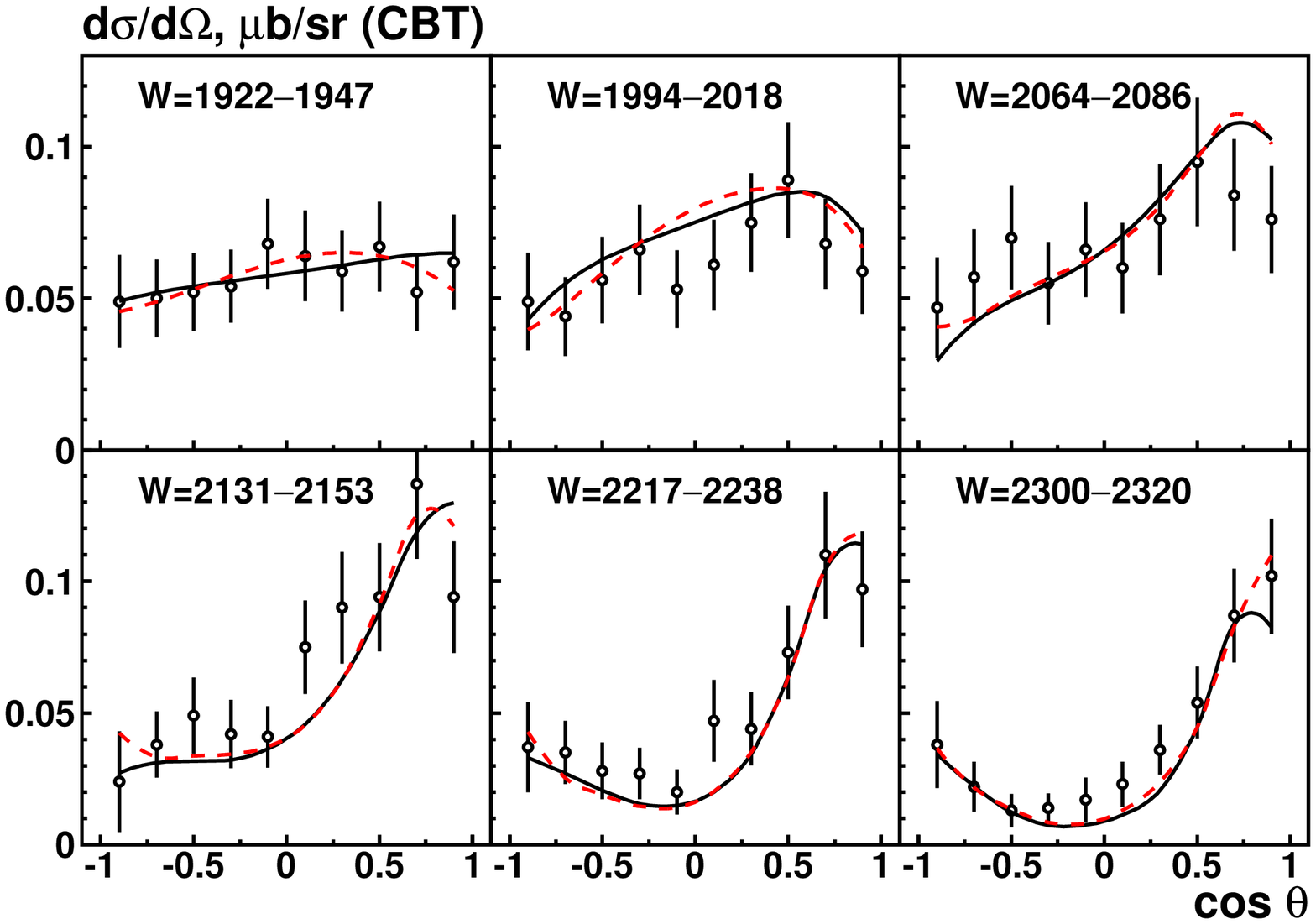}
  \end{subfigure}
  \begin{subfigure}[t]{0.48\textwidth}
    \includegraphics[width=\textwidth]{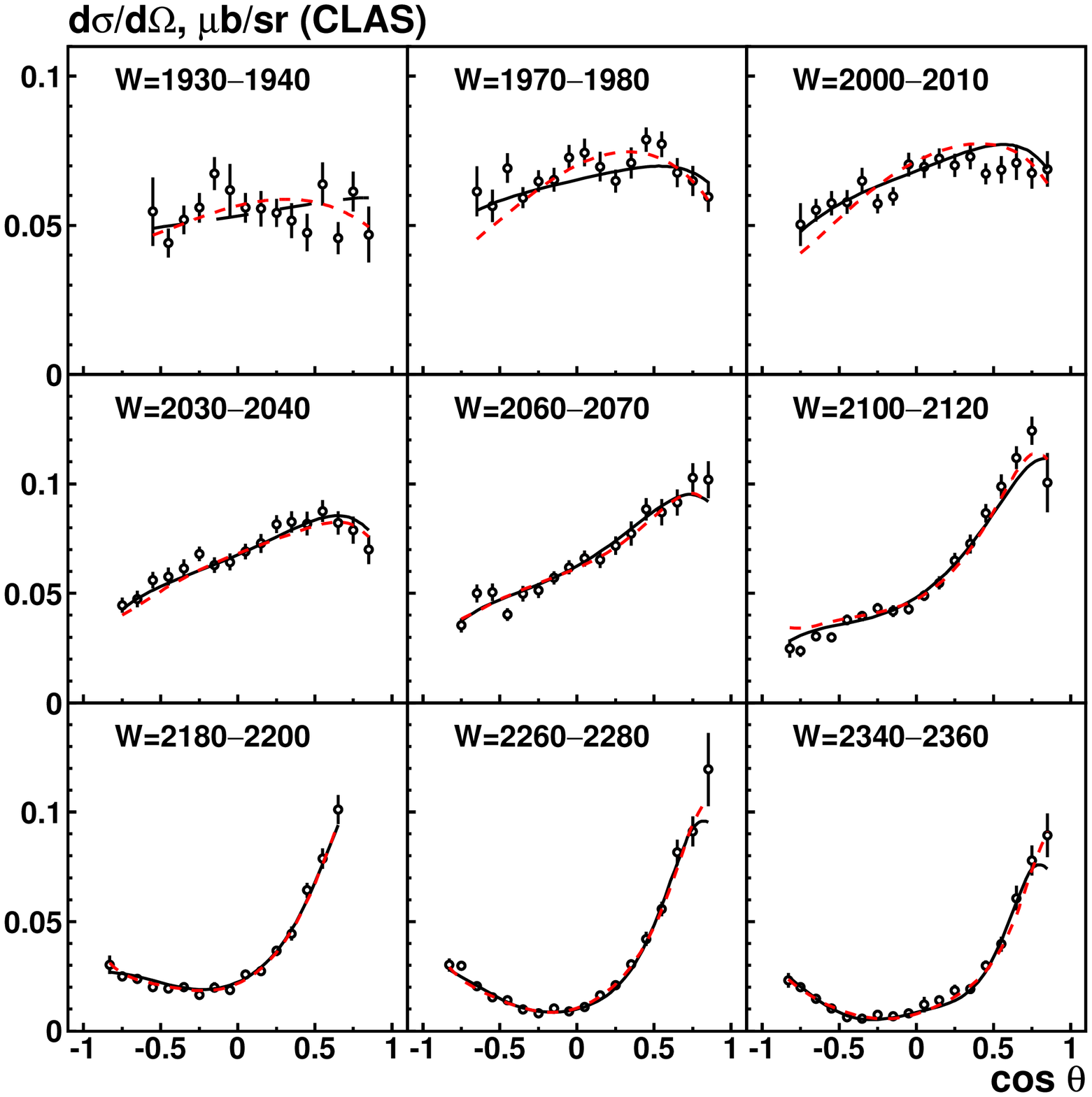}
  \end{subfigure}
  \caption{\label{fit_diff}(Color online) Selected data for the $\gamma p\to \eta' p$
    differential cross section from CBELSA/TAPS~\cite{Crede:2009zzb} (top) and
from CLAS~\cite{Williams:2009yj} (bottom).  Fit 1 is shown by solid (black)
curves and fit 2 by dashed (red) curves. The PWA-curves shown in comparison to
the CLAS data are scaled by an energy and angle independent global scaling
factor of 0.9.
}
\end{figure}
CLAS and CBELSA/TAPS differential cross sections. In the fits, the CLAS differential cross
sections were multiplied with a scaling factor $1/f$; the fit returned a scaling
factor 0.9. Thus, the curves shown for the CLAS differential cross section are multiplied
with 0.9. Likewise, the systematic uncertainty of 6\% for the beam asymmetry data from
CLAS~\cite{Collins:2017sgu} was incorporated, and the predicted beam asymmetry from the
fit was found to be larger by 6\% than the data. Beyond these three experiments, most
existing data on pion- and photo-induced reactions leading to two or three particles in
the final state were included in the fit; the dataset is described more fully in
Refs.~\cite{Anisovich:2011fc,Gutz:2014wit,Sokhoyan:2015fra}.

\begin{figure*}
 \center
\includegraphics[width=0.8\textwidth]{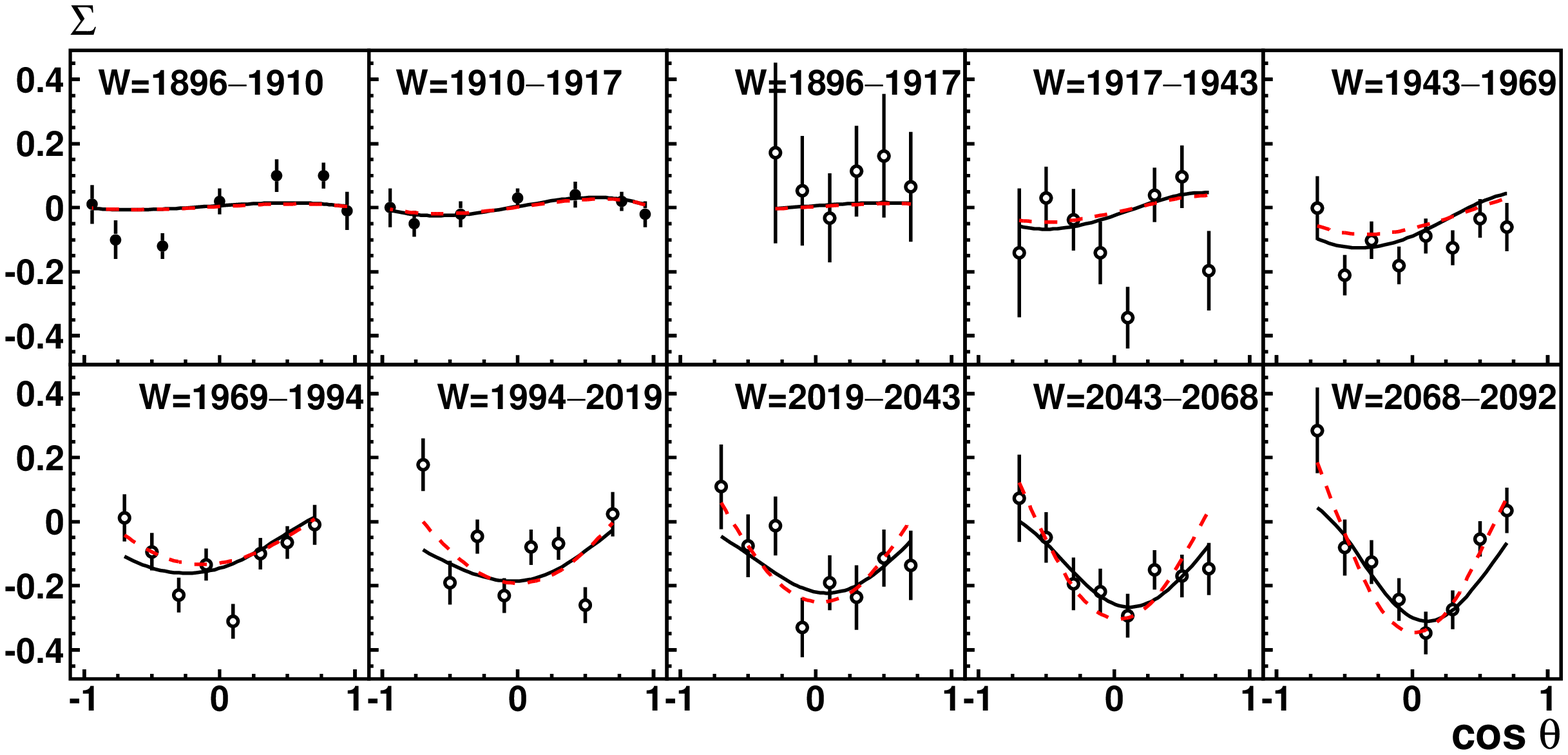}
\caption{\label{fit_sigma}(Color online) The beam asymmetry $\Sigma$
for the reaction $\gamma p\to \eta' p$. Shown are data from
GRAAL~\cite{Sandri:2014nqz} (first two subfigures) and recent data
from CLAS~\cite{Collins:2017sgu} (next eight subfigures). Also shown
are the results of each of the two partial-wave solutions discussed
in the text. Fit 1 is shown by solid (black) curves and Fit 2 with
dashed (red) curves. The PWA-curves for the CLAS beam
asymmetries are scaled by a factor 0.94. }
\end{figure*}

As might be expected when many parameters are involved, the fits of the formalism to this
dataset did not converge to a unique minimum with a single set of parameters. In the fits,
we studied the impact of different choices of high-mass resonances. The properties of
most resonances remained stable. For the reaction $\gamma p\to \eta^\prime p$, two distinct
classes of minima were found with a very similar fit quality (expressed as a total $\chi^2$).
Table~\ref{two_sol} gives the number of data points, the total $\chi^2/N_{\rm data}$, and
the breakdown of the individual contributions. Figure \ref{fit_diff} shows the differential
cross sections from the CBELSA/TAPS and CLAS experiments for selected mass bins, indicating
the general quality of the fit obtained. In general, the description of the data is
very good. At high masses, the CBELSA/TAPS differential cross section data are not fully compatible with the CLAS data; instead the former results tend to be larger. Nevertheless, both datasets can be fit simultaneously when the relative normalization is allowed to adjust the data within the stated systematic uncertainties. The resulting fit $\chi^2$ values are excellent for both datasets.

Figure~\ref{fit_sigma} shows data for the beam asymmetry $\Sigma$
from the GRAAL~\cite{Sandri:2014nqz} and the CLAS
\cite{Collins:2017sgu} experiment. GRAAL reported $\Sigma$ for two
mass bins just above the threshold. Surprisingly, the photon beam
asymmetry is already considerably strong just 7\,MeV above the
threshold and then weaker at 16\,MeV above the threshold. This rapid
change in magnitude points to large interferences among multiple
resonances below the $\eta^\prime$ production threshold.
The GRAAL beam asymmetry data could not be fit, however, unless solutions were accepted
where other data were badly described and/or where some $N^*$ had very large
$N^*\to N\eta^\prime$ coupling constants. In the end, we accepted that the GRAAL
data at 1903\,MeV could not be fully reproduced.

The CLAS data on the beam asymmetry
are also shown in Fig.~\ref{fit_sigma}. Overall, the fit obtained is reasonable, though
the fit trend is smoother than the data. The disagreement between fit and data is slightly
larger than the statistical uncertainties for several points, and, as a result, the
$\chi^2$ per data point is larger than 1 (see Table~\ref{two_sol}).

\begin{figure}[t]
  \center
  \begin{subfigure}[b]{0.4\textwidth}
    \includegraphics[width=\textwidth,height=0.8\textwidth]{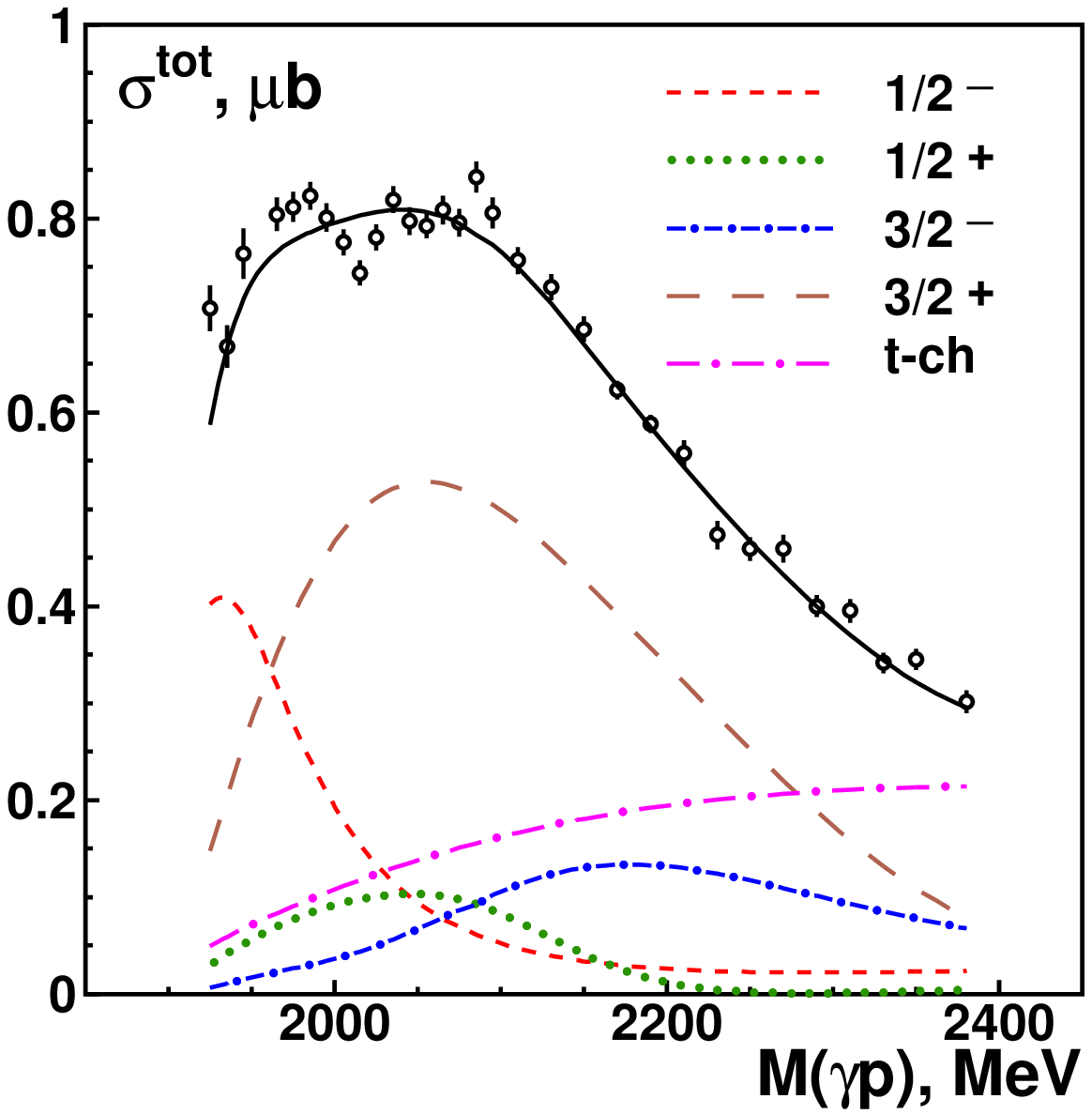}
  \end{subfigure}
  \hfill
  \begin{subfigure}[b]{0.4\textwidth}
    \includegraphics[width=\textwidth,height=0.8\textwidth]{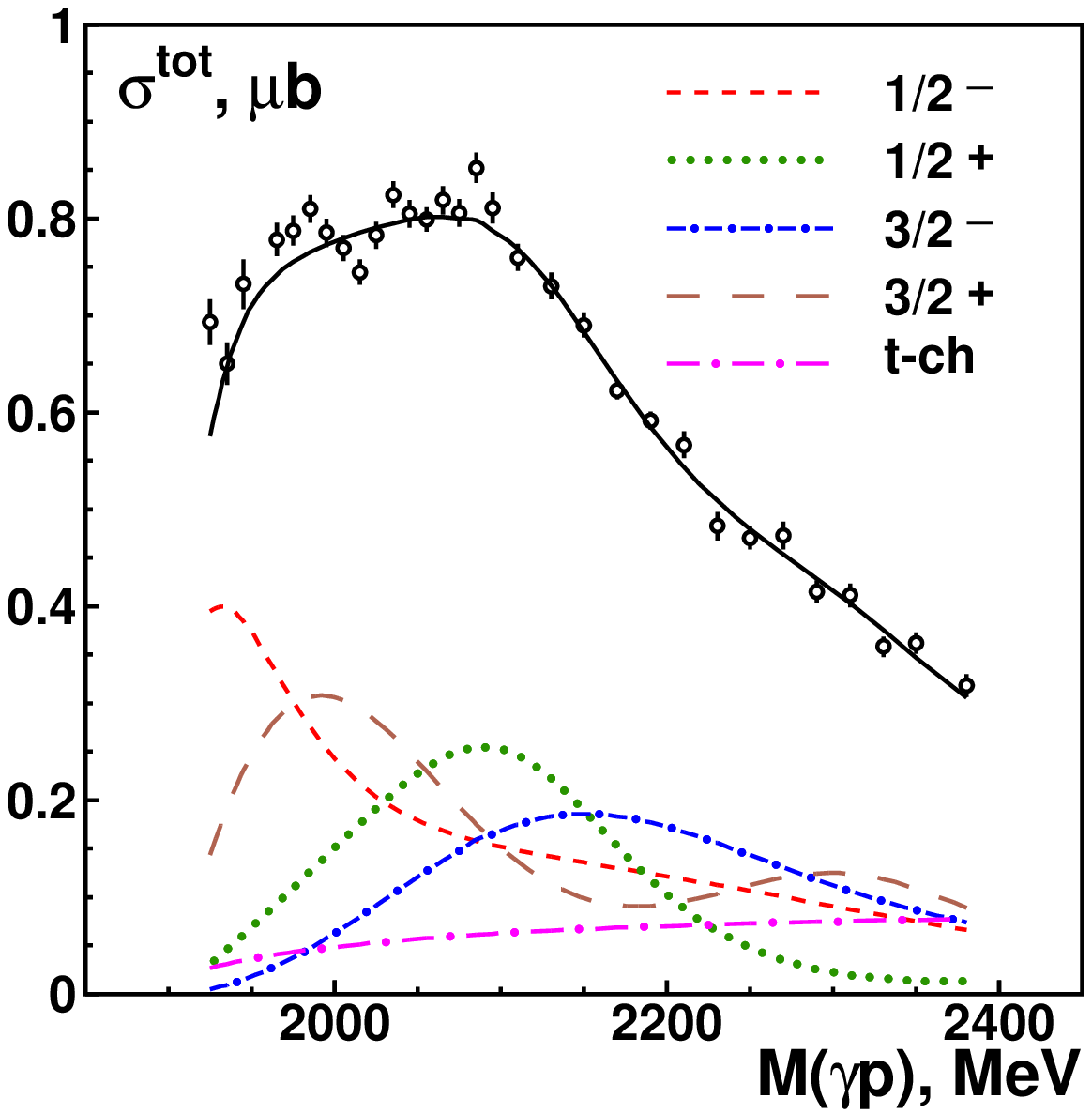}
  \end{subfigure}
  \caption{\label{fig_tot}(Color online) The total cross section for $\gamma p\to \eta^\prime p$
    and the partial wave contributions with defined spin-parity $J^P$ from the two partial-wave
    solutions described in the text. Data shown are extrapolated, as discussed in the text, from
    the CLAS data of Ref.~\cite{Williams:2009yj}. The long-dashed - dotted curve represents the
   contribution from $t$-channel exchanges. The $1/2^\pm$ and $3/2^\pm$ partial waves in the figure
        contain the contribution from $t$-channel exchange in the particular partial wave.}
\end{figure}
\begin{table}[pt]
\caption{\label{two_sol}The description of the dataset for the reaction $\gamma p
\rightarrow \eta^\prime p$ ($\chi^2/N_{\rm data}$) used in obtaining the two partial-wave
solutions discussed in the text. The remainder of the data used in the fits is described
in Ref.~\protect{\cite{Anisovich:2011fc,Gutz:2014wit,Sokhoyan:2015fra}}.\vspace{-3mm}}
\begin{center}
\renewcommand{\arraystretch}{1.4}
\footnotesize
\begin{tabular}{cccccc}
\hline\hline
&$N_{\rm data}$ &              && $\chi^2/N_{\rm data}$ & $\chi^2/N_{\rm data}$   \\[-1ex]
&               &              && Fit 1                 &  Fit 2   \\
\hline
    $d\sigma/d\Omega$  & 200 &CBT & \cite{Crede:2009zzb}   & 1.11   &  1.05   \\
    $d\sigma/d\Omega$  & 524&CLAS & \cite{Williams:2009yj} &  1.40  &  1.53   \\
    $\Sigma$           & 14 &GRAAL& \cite{Sandri:2014nqz}  &  1.49  &  1.55   \\
    $\Sigma$           & 56 &CLAS & \cite{Collins:2017sgu} &  1.78  &  1.64   \\
     total             & 794 &all &                        & 1.42   & 1.46    \\
\hline\hline
\end{tabular}
\end{center}
\renewcommand{\arraystretch}{1.0}
\end{table}

\section{Partial-wave content of the fit to $\gamma p \rightarrow \eta^\prime p$}

Having arrived at two reasonable descriptions of the differential
cross section and photon beam asymmetry data, predictions of those
fits for the total cross section can be made and the importance of
the contributions to the total cross section from various resonances
can be assessed. In comparing the CLAS data to the results of the
fit, the total cross section for the $\gamma p \rightarrow
\eta^\prime p$ reaction can be obtained by summation of the
differential cross sections over the solid angle, using the PWA fit
result for those regions of solid angle not covered by the CLAS
results.

The resulting total cross section predictions from the two fits described above are shown
in  Fig.~\ref{fig_tot}. Also shown in the figure are the most important partial-wave
contributions to the total cross section from both fits. As seen in the figure, at and
slightly above the threshold, the $N(1895)1/2^-$ resonance is seen to dominate the
reaction in both solutions. This resonance matches the characteristic energy, spin, and
parity suggested in the previous analyses of the differential cross section data mentioned
in the Introduction, where all agreed that the threshold region was governed by a
resonance with $J^P=1/2^-$ and a mass of about 1900\,MeV. The $N(1895)1/2^-$ state is a
``two star" resonance in the most recent Review of Particle Physics (RPP)~\cite{Olive:2016xmw}.
Also common to both our solutions is the strong contribution of the $N(1900)3/2^+$ excited
state, a ``three-star" resonance in the RPP. With respect to the fits reported here, however,
the contributions from $N(2100)$ ${1/2^+}$ and $N(2120){3/2^-}$ were found to be
significantly different in the two fits arrived at here. Small contributions were seen in
the fits from the $N(2060){5/2^-}$ and $N(2000){5/2^+}$ resonances, but those are not
shown in Fig.~\ref{fig_tot}.

To investigate improvements of the fit with additional resonances in the mass region above
2200\,MeV, additional poles were added to the $K$-matrix, one-by-one, with the
resulting $\chi^2$ computed.
The additional poles had spin-parities of $J^P=1/2^\pm$, $3/2^\pm$, $5/2^\pm$, $7/2^\pm$.
Although each additional state improved the quality of the fit somewhat based on $\chi^2$, no
fit with an additional contribution was obviously superior to another.
Instead, the variations in the parameters
resulting from these additions were used to estimate the model dependence of the results
for the fits.

In Table~\ref{BR}, the branching ratios deduced from these fits for $N^*\to N\eta^\prime$ decays
are presented for the six resonances used in both fits. Two of these resonances
have masses very close to the $\eta^\prime$ production threshold. Branching ratios are
usually defined at the nominal mass of a resonance and vanish identically when the threshold
is above the nominal mass. However, due to the finite width of nucleon excitations (typically
100 MeV or more), a resonance with a mass below the $\eta^\prime$ threshold may nevertheless
still decay into a final state that lies above the $\eta^\prime$ threshold.
Therefore, we use a different definition:
\begin{eqnarray}
\hspace{-8mm}BR\! =\!  \int\limits_{\rm threshold}^{\infty}\frac {ds}{\pi} \frac{f
(g_a^\alpha)^2\rho_{a}^{\alpha}(s)}{(M^{\alpha}_{\rm BW}\,^2-s)^2+ f^2
(\sum\limits_\alpha g_{a}^{\alpha}\,^2\rho_{a}^{\alpha}(s))^2}
\label{br4}
\end{eqnarray}
The Breit-Wigner mass and the parameter $f$ are adjusted so that its pole matches
the pole position of the full amplitude, $f$ is typically in the range $0.9 -1.1$.
The uncertainties in the branching ratios cover the
variations seen in these branching ratios when different primary fits were
used.

\begin{table}[pt]
\caption{\label{BR} Branching ratios (in \%) for $N^*$ resonance decays
into $N\eta^\prime$ final states based on the two partial-wave fits discussed in the text.
\vspace{-3mm}}
\renewcommand{\arraystretch}{1.3}
\begin{center}
\footnotesize
\begin{tabular}{lcccc}
\hline\hline
& \multicolumn{2}{c}{Fit 1} &\multicolumn{2}{c}{Fit 2}\\
Resonance & B.R. & $\delta(\chi^2)$ & B.R. & $\delta(\chi^2)$ \\
\hline
$N(1895){1/2^-}$   & 11\er 3 &  74 & 14\er 5 &  90 \\
$N(2100){1/2^+}$   &  8\er 2 & 105 &  7\er 2 & 150 \\
$N(2120){3/2^-}$   &  3\er 1 & 123 &  4\er 2 & 242 \\
$N(1900){3/2^+}$   &  6\er 2 & 137 &  6\er 2 & 101 \\
$N(2060){5/2^-}$   &  $<$1   &   3 &  $<$1   &   6 \\
$N(2000){5/2^+}$   &  $<$1   &   2 &  2\er 1 &  11 \\
\hline\hline
\end{tabular}
\end{center}
\end{table}

Also given in Table~\ref{BR} are the changes in $\chi^2_{\rm tot}$
when one of the resonances is omitted from the fit. The uncertainties in the branching ratios were
estimated from the spread of results obtained by adding one of the high-mass resonances.
In most cases, the uncertainties  have sizes which are similar to the
differences between Fit 1 and Fit 2.
We therefore combined these differences with the uncertainties in Table~\ref{BR} and find the
final results which are collected in Table~\ref{br-final}.

\begin{table*}
\caption{\label{br-final}Results on the four resonances with
significant $N\eta'$ decay branchings ratios. The masses and widths
$M_{\rm pole}$,$\Gamma_{\rm pole}$ are given in MeV, the moduli of
the helicity couplings in $10^{-3}$\,GeV$^{-1/2}$. Other branching ratios are given for
comparison.  
 $\Delta\pi$ denotes $\Delta(1232)3/2^+\pi$ and $N^*\pi$, $N(1520)3/2^-\pi$. \vspace{-3mm} }
\renewcommand{\arraystretch}{1.5}
\begin{center}
\begin{tabular}{lccccccccccc}
\hline\hline &$M_{\rm pole}$&$\Gamma_{\rm pole}$&
$A_{1/2}$&$A_{3/2}$
&\multicolumn{7}{c}{Branching ratios (in \%) for $N^*\to $}\\
                 &&&phase &phase
&$N\eta^\prime $&$N\pi $&$N\eta $&$N\omega$&$\Delta\pi$&$N^*\pi$&$N\sigma$\\ \hline
$N(1895){1/2^-}$ & 1907\er10 & $100^{+40}_{-10}$ &-15\er 6 && 13\er 5&2.5\er1.5&10\er5&28\er12&7\er4&-&-\\[-2.5ex]
                 &            &           &\tiny -(35\er 35)°&~& \\[-1.5ex]
$N(2100){1/2^+}$ & 2100\er 30 & 280\er 35 & -6\er 4&12\er 5& 8\er 3&16\er5 &25\er10&15\er10&10\er4&30\er4&20\er6\\[-2.5ex]
                 &            &           &\tiny (10\er 25)°&\tiny (70\er 30)°& \\[-1.5ex]
$N(2120){3/2^-}$ & 2115\er 40 & 345\er 35 &130\er 45 & 160\er 60& 4\er 2&5\er3&3\er2&12\er 8&70\er23&&11\er4\\[-2.5ex]                 &            &           &\tiny -(40\er 25)°&\tiny -(30\er 15)°& \\[-1.5ex]
$N(1900){3/2^+}$ & 1910\er 30 & 280\er 50 &26\er14 &-70\er 30&6\er 2&3\er2&3\er1&15\er9&50\er15&15\er8&4\er3\\[-2.5ex]
                 &            &           &\tiny (60\er 35)°&\tiny (70\er 50)°& \\
\hline\hline
\end{tabular}\vspace{-3mm}
\end{center}
\end{table*}

In Table~\ref{br-final} we also present pole positions, helicity
couplings, and other branching ratios which are taken from
\cite{Sokhoyan:2015fra,Denisenko:2016ugz, Muller:2017tbd}.
The branching ratios for decays into $K\Lambda$ and $K\Sigma$ will be presented
in a forthcoming publication. In two cases, the sum of the observed
branching ratios exceeds 1 but is still compatible with 1 when the
large errors are considered.

The uncertainties in the branching ratios in Table~\ref{br-final} are ranges
within which a fit solution is possible.
While these uncertainties are relatively large, the four resonances do
have statistically significant branches to $N\eta^\prime$ based on the analysis reported here.
The statistical evidence for the $\eta^\prime$ decay mode should be judged from
the change in $\chi^2$ when a resonance is removed (see Table~\ref{BR})
and not from the total uncertainties in Table~\ref{br-final}.

The most recent RPP does not report any observed branching ratios for decays
into $N\eta^\prime$. This analysis suggests that statistically
significant $N\eta^\prime$ branches exist for four of the six states indicated in
Table~\ref{BR}. If confirmed, these branches represents a useful test for any model
for the decays of those states based on, for example, a quark description of the nucleon,
especially when combined with similar data for $\eta$ photoproduction on the nucleon.

\begin{figure}[h!]
\center
\includegraphics[width=0.46\textwidth]{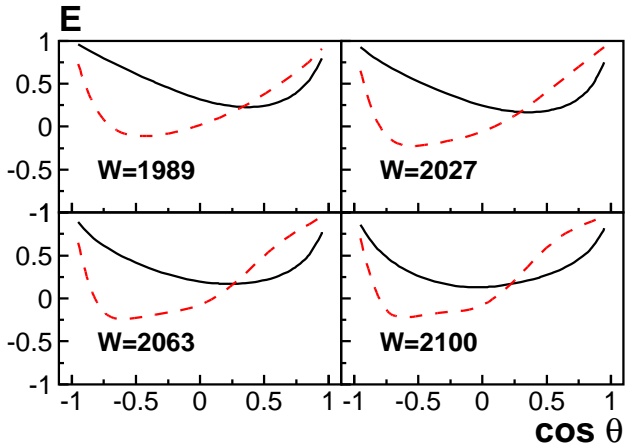}
\caption{\label{fig_E}(Color online) Prediction of the helicity asymmetry $E$ for Fit 1, solid (black) curve and for Fit 2, dashed (red) curve.
}
\end{figure}

The ambiguities in the parameters and resonances observed in the partial-wave analyses
conducted here might be resolved by further data on $\eta^\prime$ photoproduction, as
well as additional data on spin observables for the photoproduction of other mesons
from the nucleon. With respect to $\eta^\prime$, the CBELSA/ TAPS collaboration
has presented preliminary results on the helicity difference $E=(\sigma_{1/2} -
\sigma_{3/2})/(\sigma_{1/2} + \sigma_{3/2})$ for the $\gamma p\to \eta^\prime p$
reaction \cite{Afzal:2012fna}. Since those data are preliminary, those data were not
yet included in the present fits. However, when finalized, those results may help to
discriminate between the two fit solutions and to better constrain the parameters and
branching ratios obtained in this work. With that in mind, Figure~\ref{fig_E} shows
our prediction for the helicity asymmetry $E$ for the $\gamma p\to \eta^\prime p$
reaction based on the two partial-wave solutions found in this work.

\section{Conclusion}
We have included new data on the reaction $\gamma p\to \eta^\prime p$ in a fit to
the BnGa database for meson photoproduction. A partial-wave analysis identifies the
leading resonance contributions and provides additional support for the existence
of $N(1895)1/2^-$, $N(1900)3/2^+$, $N(2100)1/2^+$, and $N(2120)3/2^-$. When analyzed
in terms of these resonances, statistically significant branching ratios for their
decays into $N\eta^\prime$ were determined. In the most recent PDG summary, no such
branches are indicated as having been seen. Additional data on spin observables
should help pin down the issue of whether these branches to $N\eta^\prime$
are present for the resonances of Table~\ref{BR}, which, in turn, will help test models
of the nucleon.

\section{Acknowledgments}
This work was supported by the Deutsche Forschungsgemeinschaft (SFB/TR110) and by the
RSF grant 16-12-10267. Work at Arizona State University was supported by the U.~S.~National
Science Foundation under award PHY-1306737. This material is in part based upon work
supported by the U.S. Department of Energy, Office of Science, Office of Nuclear Physics
under contract DE-AC05-06OR23177.

\end{document}